\documentclass[]{spie}

\usepackage{graphicx}
\usepackage{multirow}
\newcommand{\MI}{\mathrm{MI}}
\newcommand{\SNR}{\mathrm{SNR}}
\usepackage{amsfonts}
\usepackage{amsmath}
\usepackage{amsmath,bm} 
 
\usepackage{amsmath,amsfonts,amssymb}
\usepackage{graphicx}
\usepackage[colorlinks=true, allcolors=blue]{hyperref}
\usepackage{xcolor, soul}

\title{Anatomy-based quality metric of diffusion-weighted MRI data for accurate derivation of muscle fiber orientation}

\author[a]{Nadya Shusharina}
\author[b]{Xiaofeng Liu}
\author[c]{Evangelia Kaza}
\author[c]{Miranda Lam}
\author[d]{Stephan Maier}
\author[e]{Jonghye Woo}

\affil[a]{Dept. of Radiation Oncology, Massachusetts General Hospital and Harvard Medical School, Boston, MA 02114, USA}
\affil[b]{Dept. of Radiology and Biomedical Imaging, Yale University, New Heaven, CT 06519, USA}
\affil[c]{Dept. of Radiation Oncology, Brigham and Women's Hospital, Dana-Farber Cancer Institute and Harvard Medical School, Boston, MA 02115, USA}
\affil[d]{Dept. of Radiology, Brigham and Women's Hospital and Harvard Medical School, Boston, MA 02115, USA}
\affil[e]{Gordon Center for Medical Imaging, Massachusetts
General Hospital and Harvard Medical School, Boston, MA 02114 USA}
\authorinfo{Send correspondence to Nadya Shusharina\\
E-mail: nshusharina@mgh.harvard.edu}

\begin{document}

\maketitle              % typeset the header of the contribution
\begin{abstract}
Diffusion-weighted MRI (DW-MRI) is used to quantitatively characterize the microscopic structure of soft tissue due to the anisotropic diffusion of water in muscle. Applications such as fiber tractography or modeling of tumor spread in soft tissue require precise detection of muscle fiber orientation, which is derived from the principal eigenvector of the diffusion tensor. For clinical applications, high image quality and high signal-to-noise ratio (SNR) of DW-MRI for fiber orientation must be balanced with an appropriate scan duration. Muscles with known structural heterogeneity, e.g. bipennate muscles such as the thigh rectus femoris, provide a natural quality benchmark to determine fiber orientation at different scan parameters. Here, we analyze DW-MR images of the thigh of a healthy volunteer at different SNRs and use PCA to identify subsets of voxels with different directions of diffusion tensor eigenvectors. We propose to use the mixing index of spatial co-localization of the clustered eigenvectors as a quality metric for fiber orientation detection. Comparing acquisitions at different SNRs, we find that high SNR results in a low mixing index, reflecting a clear separation of the two compartments of the bipennate muscle on either side of the central tendon. 
Because the mixing index allows joint estimation of spatial and directional noise in DW-MRI as a single parameter, it will allow future quantitative optimization of DW-MRI protocols for soft tissue.

\end{abstract}
\section{Introduction}
In managing soft tissue sarcoma (STS), it is critical to determine the extent of microscopic tumor spread, which is not visible using currently available in-vivo medical imaging.  It is delineated on a computed tomography (CT) scan by varying degrees of longitudinal and transverse expansion of the radiographically visible tumor mass. The accuracy of manual delineation from 2D views is constrained by the inadequate visual representation of the anatomy's 3D structure. Consequently, the delineation of the extent remains a matter of clinical judgment, subject to variability in expert opinion \cite{Vinod2016,Genovesi2014}. This may result in a higher risk of treatment failure. 

Although microscopic tumor spread is not visible in conventional medical imaging, it can be modeled using diffusion-weighted MRI (DW-MRI). It has been confirmed that tumor cells invade soft tissues by occupying the 3D spaces defined by tissue interfaces \cite{Weigelin2012}.  DW-MRI has proven useful for studying the microscopic structure of soft tissues due to the anisotropic water diffusion found in muscle \cite{Damon2016}, and therefore DW-MRI data can be used to derive tissue directionality. With DW-MRI, the extent of microscopic spread can be defined as the distance from the radiographically visible tumor along the muscle fibers \cite{Shusharina2022}.  

The principal eigenvector of the diffusion tensor derived from DW-MRI data coincides with the fiber directionality.  For applications that require precise detection of the fibers, such as models of tumor spread, the consistency of the principal eigenvector is of major importance. Previous studies have addressed the uncertainties of DW-MRI-derived parameters as a function of image acquisition setup \cite{Froeling2010,Rockel2016,Monte2020,Shusharina2023}. These studies agreed that the level of signal-to-noise ratio (SNR) is one of the most important factors in determining the accuracy of diffusion tensor parameters. Because muscles have a long T$_1$ and a very short T$_2$, achieving a high SNR requires long acquisition times, which is unfavorable for scanning cancer patients. Optimization of the protocol for image acquisition combining acceptable scan time and high image quality is a challenging task and requires well-defined metrics for optimization.

In this work, we introduce mixing index of spatial co-localization of voxels with distinct principle eigenvector direction as an accuracy measure for optimal SNR. 
Low values of the mixing index ensure acceptable image data quality for the application of the soft tissue tumor spread model. To prototype a possible fully automated imaging workflow, we have compared U-Net based and manual muscle segmentation and their impact on fiber direction analysis.

\section{Materials and Methods}
\subsection{Imaging} We acquired morphological and DW-MR images of one thigh of a healthy volunteer on a 3T MRI system. A T$_1$-weighted MR sequence was acquired with the repetition time (TR) of 4.6 ms, and the echo time (TE) of 2.77 ms, and 128 axial slices of 2 mm thickness and 1 mm$^2$ in-plane resolution. The DW-MRI acquisition consisted of two $b_0=50$ s/mm$^2$ images and 12 diffusion-weighted images with $b=400$ s/mm$^2$ using 12 gradient directions. The TR was set to 8,000 ms and TE to 45 ms. Images were acquired in the axial plane with a slice thickness of 6 mm and 38 slices. Two DW-MRI scans were performed, one with a low planar resolution having a pixel size of $3\times3$ mm$^2$, and the other with a high planar resolution with a pixel size of $1.3\times1.3$ mm$^2$. 

\subsection{Structure Labeling} 
\begin{figure}[th]
\includegraphics[width=\textwidth]{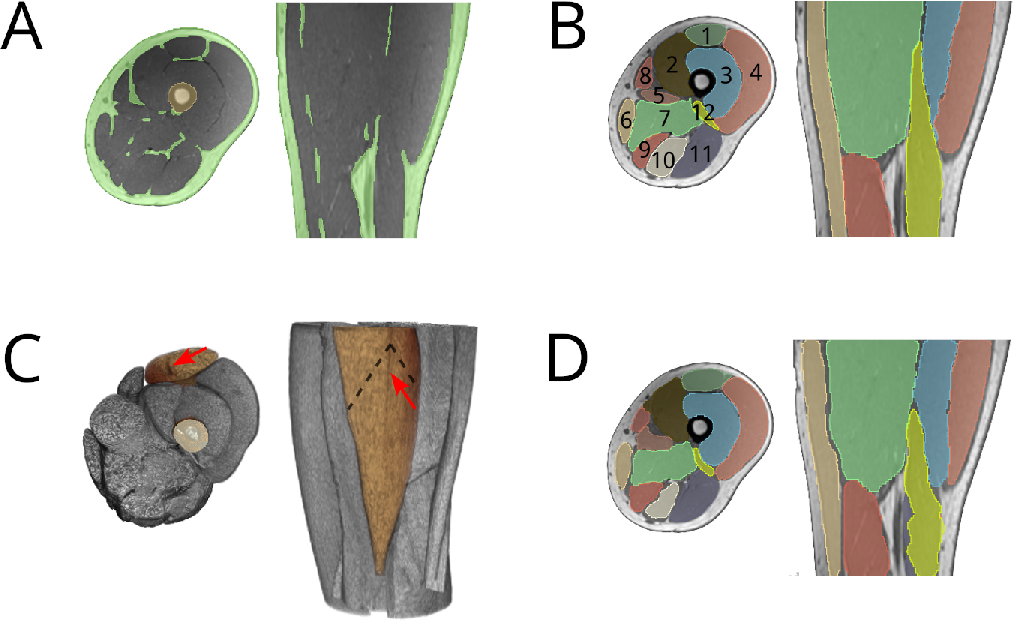}
\caption{T$_1$-weighted MR image of the thigh, axial and coronal views. Panel A: Subcutaneous and intramuscular fat (green) and femur (orange) segmented by image intensity thresholding. Panel B: Manually segmented individual muscles, 1 (RF), 2 (VM), 3 (VI), 4 (VL), 5 (AL), 6 GRA), 7 (AM), 8  (SAR), 9  (SM), 10  (ST), 11  (BFL), 12 (BFS).  Panel C: 3D rendering of the muscles. The rectus femoris muscle is shown in dark orange. The central tendon is visible in both views (red arrow). The fiber orientations in both parts of the muscle are  shown schematically as dashed lines.  Panel D: Results of deep learning based auto-segmentation of individual muscles.} \label{segmentation}
\end{figure}

The femoral bone and fat (subcutaneous and intramuscular) were segmented on the T$_1$-weighted MR image by intensity thresholding (Fig.~\ref{segmentation}, panel A). For automated muscle segmentation, we used the convolutional neural network model with 2D U-net architecture by Ronneberger et al. \cite{Ronneberger2015} using TensorFlow, trained as described in \cite{Shusharina2022}, using cross-entropy loss for supervised learning as in \cite{Liu2021}. The network was trained using multi-channel input, T$_1$-weighted MR volumes, apparent diffusion coefficient (ADC) volumes, and fractional anisotropy (FA) volumes (three channels). Only the T$_1$ image is required for inference. After applying the model to define the muscle masks, the following post-processing was performed. The overlap between the muscle masks and the bone and fat was removed by prioritizing the thresholded tissue. Then, small island removal and dilation-erosion standard smoothing operations were applied. 

Twelve individual muscles, including the rectus femoris (RF), vastus medialis (VM), vastus intermedius (VI), vastus lateralis (VL), adductor longus (AL), gracilis (GRA), adductor magnus (AM), sartorius (SAR), semimembranosus (SM), semitendinosus (ST), biceps femoris long head (BFL), and biceps femoris short head (BFS), were identified manually, as shown in Fig.~\ref{segmentation}, panel B and by automated segmentation as shown in Fig.~\ref{segmentation}, panel D. Among the 12 muscles, one muscle, the rectus femoris, is bipennate with the fibers divided into two parts between the two sides of the central tendon at different angles to the longitudinal axis of the muscle (pennate angle), as shown in Fig.~\ref{segmentation}, panel C. 

\subsection{Signal-to-Noise Ratio}
The SNR of DW-MR images was computed within each individual muscle as the ratio of the mean signal, $S_{mean}$, and the standard deviation of the underlying Gaussian noise, $\sigma$:
\begin{equation}
\SNR=\frac{S_{mean}}{\sigma}.
\label{snr}
\end{equation}
The noise, denoted as $\sigma$, was calculated as the local noise variance within each voxel. This calculation was performed using the local PCA-based algorithm~\cite{Manjon2013} implemented in the Python library DIPY \cite{Dipy}. The average magnitude of the DW-MRI signal in each voxel was calculated by first averaging the data over 12 repetitions for each gradient direction. The signal magnitude, $S_{mean}$, was defined as the root mean square amplitude across all gradient directions for each voxel within each muscle.
\subsection{Diffusion Tensor and Directional Variability} The diffusion-weighted series were resampled to an isotropic voxel size equal to the planar resolution of the acquired DW-MRI image. The diffusion tensor was reconstructed using DIPY with the tensor model of Basser et al.~\cite{Basser1994}. Directional variability was quantified as the voxel-wise angular deviation from the principal eigenvector, averaged over all the voxels within a muscle. To remove data degeneracy due to up and down symmetries in the direction of the principal eigenvector, two eigenvector clusters were detected using $k$-means, and the direction of the vector in one of the two clusters was reversed. For each muscle, the mean vector was then calculated. A voxel-wise angle between mirrored eigenvectors and mean vector was then computed. Histograms of voxel-wise deviations were plotted for each muscle and the mode of the histogram, referred to as directional variability (DV), was reported.

\subsection{PCA on the Principal Eigenvectors of the Diffusion Tensor}To separate the two compartments of a bipennate muscle on either side of its central tendon, PCA analysis was performed in the space of nine coordinates of the three eigenvectors of the voxel-wise diffusion tensor within the rectus femoris. Dimensionality reduction was performed using the three largest components. Then, $k$-means clustering was used to identify four classes of eigenvectors, with two main classes being mirrored components of the principal eigenvector directions. Within one of the main classes, three classes were identified, two classes of principal eigenvectors within each of the compartments oriented along fibers with different pennate angles, and one class of principal eigenvectors with random orientation.

\subsection{Mixing Measurement}To characterize the quality of the imaging data for identifying the two compartments in the bipennate muscle, we calculated the degree of separation of the voxels belonging to different classes of principal eignevector. We used a nearest-neighbor mixing index, which is widely used to characterize the homogeneity of the solid particle mixture. On the three-dimensional voxel grid, the degree of mixing is determined by calculating the proportion of nearest voxels that belong to a class other than the class of the central voxel. The mixing index ($\MI$) for k classes is defined by the following equation:
\begin{equation}
    \MI=\frac{1}{N_{total}}\sum_k\left[\frac{N_{total}}{N_{total}-N_k}\sum_{i=1}^{N_K}\frac{p_i-n_i}{p_i}\right],
    \label{MI}
\end{equation}
where $N_{total}$ is the number of voxels in the rectus femoris muscle, $N_k$ is the number of voxels of class $k$, $n_i$ is the number of nearest neighbors of voxel $i$ that belong to a class same as the class of that voxel, and $p_i$ is the number of the nearest neighbors of voxel $i$. $\MI$ approaches 0 in the case of complete separation and $\MI=1$ corresponds to a homogeneous mixture.

\section{Experiments and Results}
\subsection{Segmentation Model}Fig.~\ref{segmentation}, panel C shows the auto-segmentation  of the 12 individual muscles on the T1-weighted MR image, and Table~\ref{tab1} lists the Dice similarity coefficient (DSC) between automated and manual segmentations, demonstrating a moderate to high accuracy of the U-Net model.  
\begin{table}[h]
\centering
\setlength{\tabcolsep}{5.5pt}
\renewcommand{\arraystretch}{1.2}
\caption{Segmentation accuracy: Dice similarity coefficient DSC for 12 thigh muscles.}\label{tab1}
\begin{tabular}{|c|c|c|c|c|c|c|c|c|c|c|c|}
\hline
RF   &   VM & VI   & VL   & AL   & GRA  & AM   & SAR  & SM   & ST   & BFL  & BFS\\ \hline
0.80 & 0.85 & 0.85 & 0.88 & 0.79 & 0.64 & 0.85 & 0.59 & 0.77 & 0.79 & 0.81 & 0.70 \\ \hline
\end{tabular}
\end{table}

\subsection{Directional Variability} Two consecutive scans of the thigh were performed using different voxel grids. For a coarse grid with a lateral resolution of $3\times3$ mm$^2$, the SNR calculated using Eq.~(\ref{snr}) varied across different muscles, ranging from 49 to 90 (mean: $65\pm13$). The variability of the principal eigenvector direction ranged from $6^{\circ}$ to $14^{\circ}$ (mean $10^{\circ}\pm3^{\circ}$). For the finer grid with the lateral resolution of $1.3\times1.3$ mm$^2$, the SNR decreased with the range from 24 to 43 (mean $33\pm6$) with directional variability ranging from $8^{\circ}$ to $20^{\circ}$ (mean $12^{\circ}\pm4^{\circ}$). 

\begin{figure}[h]
\includegraphics[width=\textwidth]{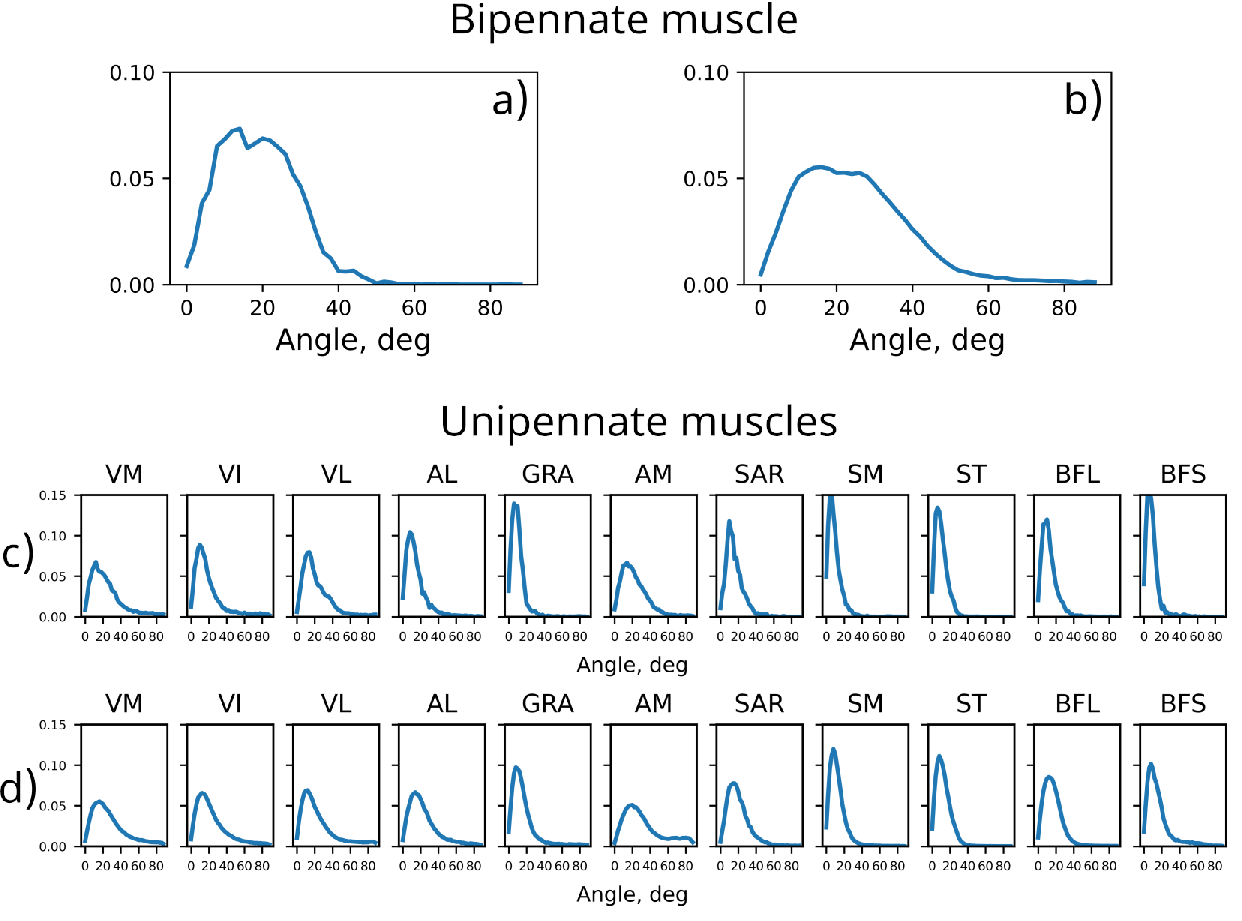}
\caption{The histograms of directional variability of the principal eigenvector plotted for the bipennate rectus femoris muscle (top row) scanned with a pixel size of $3\times3$ mm$^2$ (a) and $1.3\times 1.3$ mm$^2$ (b). The histograms for the other 11 unipennate muscles are shown for a case of pixel size of $3\times3$ mm$^2$ (c) and $1.3\times1.3$ mm$^2$ (d).} \label{angle_dev}
\end{figure}

For the bipennate rectus femoris muscle, the SNR was 89.7 at low resolution and 43.9 at high resolution. Since there are two pennate angles, bimodal distribution of the histogram of the principal eigenvector variability should be expected. Indeed,  for the low resolution scans with higher SNR,  two peaks of the histogram are pronounced (Fig.~\ref{angle_dev}(a)).   At the high resolution and the correspondingly lower SNR, the peaks merge and the bimodality disappears (Fig.~\ref{angle_dev}(b)). For comparison, unimodal histograms for the other 11 unipennate muscles are also shown for low (Fig.~\ref{angle_dev}(c)) and high (Fig.~\ref{angle_dev}(d)) resolution.

\subsection{Eigenvector PCA and Mixing Index} 
\begin{figure}[t]
\includegraphics[width=\textwidth]{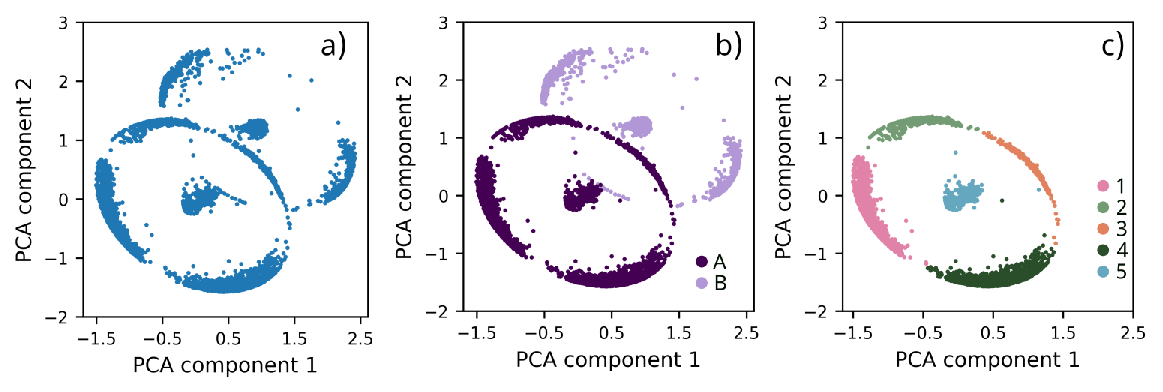}
\caption{PCA results for the 9-dimensional dataset of eigenvector coordinates in the rectus femoris muscle. Dimensionality was reduced to three components and first two components plotted in 2D.  (a) all points in the dataset; (b) all points partitioned into two clusters by $k$-means; (c) cluster A partitioned into 5 clusters by $k$-means.} \label{pca}
\end{figure}

\begin{figure}[t]
\centering
\includegraphics[width=12cm]{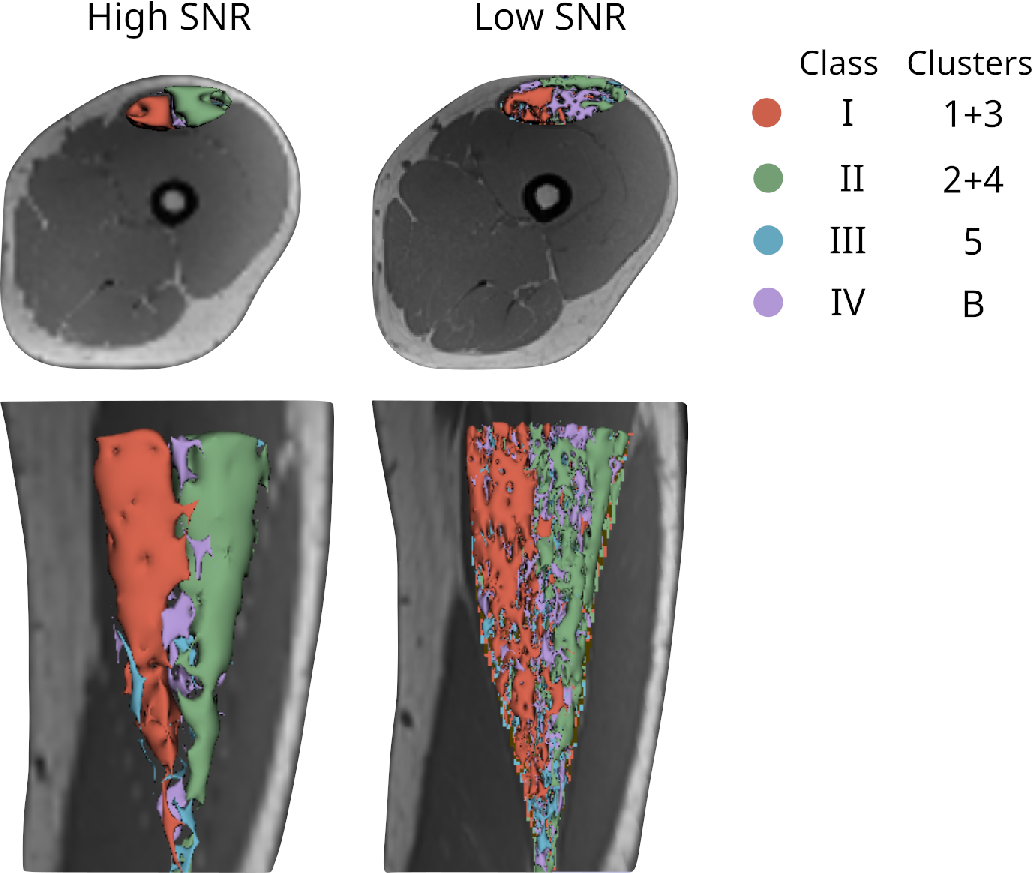}
\caption{3D labelmaps of the clustered 3-component PSA points overlaid with T$_1$ weighted MRI scan of the thigh. Two major labels (red and green) correspond to the fiber directions in the two compartment of the rectus femoris muscle. Two other labels are a random component (blue) and mirrored direction of the principal eigenvector (purple).} \label{clusters}
\end{figure}

Fig.~\ref{pca} shows the results of the PCA analysis for the eigenvector coordinates within the rectus femoris scanned at low resolution (high SNR). The results are plotted as 2D scatterplots with the first and second PCA components on the x and y axes, respectively, Fig.~\ref{pca}(a). Using the first 3 PCA components, all of the points can be clearly divided into two clusters using the k-means method, with two clusters representing the direction and the mirrored direction of the principal eigenvector (clusters A and B in Fig.~\ref{pca}(b)). There are 4031 points in cluster A and 1017 points in cluster B.  Voxels belonging to clusters A and B are randomly distributed in space (data not shown). Cluster A can be further divided into five clusters (1 through 5 in Fig.~\ref{pca}(c)). The points of the five clusters and of cluster B were translated into the labelmap with six labels. The labels of cluster 1 were combined with those of cluster 3, and the same was done for clusters 2 and 4. The final label map with four labels is shown in Fig.~\ref{clusters} (axial view and coronal view). The same workflow was used to create the labelmap for the low SNR case, as shown in Fig.~\ref{clusters}. 
	
To characterize separation of the principal eigenvector directions between the two compartments, the mixing index $\MI$ was calculated for the four-component mixture using Eq.~(\ref{MI}). As the two components belonging to class III and IV were randomly distributed in the muscle, we excluded these two classes and calculated MI for the mixture of components of class I and II.  For the high SNR case the $\MI=0.16$ and for the low SNR case $\MI=0.22$. 
The PCA analysis was repeated for the rectus femoris muscle delineated by automated segmentation, DSC$=0.80$. While the pattern was generally similar to Fig.~\ref{pca}, clear separation of the clusters was prevented by voxels belonging to other surrounding muscles. We conclude that for combining automated segmentation with fiber directionality analysis, DSC of 0.80 is not sufficient.

\section{Discussion and Conclusion}
Diffusion tensor parameters are sensitive to noise levels, and long acquisitions are not possible in clinical settings, so there is a trade-off between image data quality and patient comfort. In this work, we proposed and calculated the mixing index, a novel metric for optimizing DW-MRI acquisition of skeletal muscle. Our proposed criterion is based on the accuracy of identifying the mutual arrangement of muscle fibers within a bipennate muscle to define image quality. This is critical for applications in the treatment of cancer patients, as it defines a direction of microscopic tumor progression. Using the bipennate rectus femoris muscle as an example, we have shown that the mixing index is a function of SNR. The mixing index jointly characterizes the angular and spatial uncertainties of DW-MR image acquisition. It provides a convenient scalar metric to optimize DW-MR parameters and SNR to achieve acceptable data quality for fiber directionality modeling. We believe this is the first approach to directly link DW-MR image quality with anatomical quantification. Future developments will determine the exact relationship between mixing index, SNR, and acquisition parameters, and bring this technique closer to clinical practice.

\end{document}